\documentclass[aps,prl,amssymb,superscriptaddress,nofootinbib,twocolumn,preprintnumbers]{revtex4-1}
\usepackage[utf8]{inputenc}
\usepackage[english]{babel}
\usepackage{amsmath}
\usepackage{amsfonts}
\usepackage{amssymb}
\usepackage{amsthm}
\usepackage{color}
\usepackage{graphicx}
\usepackage[force]{feynmp-auto}
\usepackage{feynmp}
\usepackage[hidelinks]{hyperref}

\newcommand\bw{\begin{widetext}}
\newcommand\ew{\end{widetext}}

 \def\be{\begin{equation}}
\def\ee{\end{equation}}
 \def\ba{\begin{align}}
\def\ea{\end{align}}
\def\bea{\begin{eqnarray}}
\def\eea{\end{eqnarray}}

\def\m{\mu}

\begin{document}
\title{{\bf A Path (Integral) to Scale Invariance}}
\author{Mario Herrero-Valea}
\email[]{mherrero@sissa.it}
\address{SISSA, Via Bonomea 265, 34136 Trieste, Italy and INFN Sezione di Trieste}
\address{IFPU - Institute for Fundamental Physics of the Universe \\Via Beirut 2, 34014 Trieste, Italy}

\begin{abstract}
We propose a path integral formulation for scale invariant quantum field theories. We do it by modifying the functional integration measure in such a way that the partition function is always exactly scale invariant, at the cost of having an extra determinant under the integral. In perturbation theory, this extra determinant reproduces the bottom-up procedure of scale invariant regularization, providing an order-by-order cancellation of the scale anomaly together with new non-renormalizable vertices. Our formulation here, however, goes beyond perturbation theory and it is also suitable to study non-perturbative effects. It allows to formulate a scale invariant quantum theory out of any classically invariant action.
\end{abstract}

\maketitle
%%%%%%%%%%%%%%%%%%%%%%%%%%%%%%%
%%%%%%%%%%%%%%%%%%%%%%%%%%%%%%%%
%%%%%%%%%%%%%%%%%%%%%%%%%%%%%%%%
%%%%%%%%%%%%%%%%%%%%%%%%%%%%%%%%
%%%%%%%%%%%%%%%%%%%%%%%%%%%%%%%%
%%%%%%%%%%%%%%%%%%%%%%%%%%%%%%%%
%%%%%%%%%%%%%%%%%%%%%%%%%%%%%%%%
\paragraph*{{\bf Introduction} -}
One of the most profound ordering principles of Nature is that of \emph{symmetry}. Physical systems obey conservation laws inherited from invariance of the action under certain transformations. This principle has become a basic building brick when constructing theoretical models within (Quantum) Field Theory. A very successful example of this philosophy is the Standard Model of Particle Physics (SM), built upon the combination of Poincaré Invariance and a gauge group ${\rm SU}(3)\times{\rm SU}(2)\times {\rm U}(1)$. Even more interesting is the fact that symmetries can be used as guiding principles to expand well-tested theories into the unknown within the context of Effective Field Theory (EFT).

A symmetry of great interest in several fields is \emph{scale invariance} (SI) \cite{Wetterich:2019qzx}, defined as invariance under constant dilatations of all physical scales, such as lengths or energies\footnote{The terms `scale invariance', `dilatations', and `conformal symmetry' are normally used in a interchangeable way in the literature. Here by dilatations we strictly mean constant rescaling of scales.}
\begin{align}
L \rightarrow \Omega^{-1}L,\qquad E\rightarrow \Omega E.
\end{align}

Although our Universe is clearly not scale invariant due to the existence of mass scales, there are hints that SI might be a relevant symmetry in several situations. The high-energy dynamics of the SM is one of such examples. When energies are much larger than the Electroweak scale $v\approx 246\ {\rm GeV}$, particles are effectively massless and all dimensionful couplings are suppressed, leaving an approximately SI theory. A similar situation occurs in Cosmology. Observations of the CMB power spectrum point to an early Universe which was \emph{almost} SI \cite{Aghanim:2019ame}, the symmetry just softly broken at most.

These situations suggest an interesting idea. What if the fundamental theories describing these phenomena are actually SI but this symmetry is \emph{spontaneously broken} at low energies? One can easily engineer classical extensions of both the SM \cite{Chankowski:2014fva,Salvio:2014soa} and inflationary models \cite{Rubio:2020zht,Shaposhnikov:2008xi,Gorbunov:2013dqa,Rubio:2017gty,Wetterich:1987fm} to incorporate SI by introducing a dilaton field $\sigma$, whose vacuum expectation value (vev) sets a universal scale out of which the rest of scales are born. This nice idea is however obstructed in Quantum Field Theory (QFT), due to the presence of the scale anomaly. Even if the classical theory is SI, quantum corrections will violate this symmetry through the introduction of an explicit scale -- the cut-off scale $\Lambda$. Only when the theory flows to a fixed point, SI is recovered.

A possible solution to this problem comes under the name of \emph{scale invariant regularization} (SIR). Introduced in \cite{Englert:1976ep}, it regained momentum in recent years through application to modern problems in Cosmology \cite{Armillis:2013wya,Shaposhnikov:2008xi,Shaposhnikov:2008xb,Gretsch:2013ooa,Ferreira:2018qss,Ferreira:2018itt,Ferreira:2016vsc,Shaposhnikov:2018nnm} and particle physics \cite{Ghilencea:2017yqv,Ghilencea:2016ckm}. The core idea behind SIR is to replace the cut-off $\Lambda$ by the dilaton $\sigma$ after computing loop corrections. The outcome of this procedure is then automatically SI at the given order, not containing any explicit scale. Only when SI is spontaneously broken by letting $\sigma=\langle \sigma\rangle+\delta\sigma$, we recover the standard result with a constant renormalization scale $\Lambda=\langle \sigma\rangle$ \cite{Tamarit:2013vda}, plus new suppressed interactions coming from the logarithms in the finite piece of loop corrections
\begin{align}
\log \left(\sigma\right)=\log\left(\Lambda\right)+\sum_{n=1}^{\infty} \frac{(-1)^n}{n} \left(\frac{\delta\sigma}{\Lambda}\right)^n.
\end{align}

SIR has several limitations however. First, the procedure can only be applied order by order in the loop expansion, lacking a non-perturbative definition. Second, due to the non-polynomial character of the logarithm, the theory only makes sense as a perturbative QFT in the broken phase, after expanding $\sigma$ around its vev. When written in this way, the theory is always non-renormalizable \cite{Shaposhnikov:2009nk,Mooij:2018hew}. It is then reasonable to ask -- is it possible to evade, or relax, these issues and formulate a SI QFT from first principles?

This issue is not trivial, since the only known examples of QFTs enjoying scale invariance and admitting spontanous symmetry breaking are rather exotic theories. Most of them rely on the existence of Supersymmetry (SUSY), for instance ${\cal N}=4$ Super-Yang-Mills \cite{Sohnius:1981sn}. Recently, it has been also proposed that the same might be true for `fishnet' conformal field theories \cite{Karananas:2019fox}. No other sensible SI QFT theory is known, much less any model resembling the low-energy spectrum of the SM.

In this \emph{letter} we propose a path integral formulation of SIR. We do it by defining a exactly SI path integral for any QFT which is classically SI, through a modification of the functional measure of integration. The resulting QFT is therefore equally appropriate to study both perturbative and non-perturbative effects. We will also show how the perturbative expansion  exactly matches the bottom-up procedure of SIR.
\\

\paragraph*{{\bf Classical and Quantum SI} -}
Let us start by considering the classical theory of a set of interacting fields that we denote collectively as $\phi^a$, with action
\begin{align}
S[\phi]=\int d^4x \ {\cal L}[\phi].
\end{align}
Here $a$ must be interpreted as a DeWitt super-index, condensing both space-time indices and a label for the different fields in the action. We will assume that $S[\phi]$ is classically SI, meaning invariance under
\begin{align}
x^\m\rightarrow \Omega^{-1} x^\m ,\qquad \phi^a \rightarrow \Omega^{d_{\phi^a}} \phi^a  ,
\end{align}
where $d_{\phi^ a}$ is the scaling dimension of the field $\phi^a$.

A naive consequence of SI is the absence of any dimensionful constant in the Lagrangian. However, we have observed plenty of dimensionful couplings in our Universe. Hence, SI cannot be realized at the energies that we have accessed so far. This clash can be circumvented by replacing all dimensionful couplings by powers of a dilaton $\sigma$. All scales can then be generated at low energies by a non-vanishing vev of the dilaton $\langle \sigma \rangle$ combined with a dimensionless quantity. For instance, let us consider the following potential
\begin{align}\label{eq:potential_two_fields}
V(h)=\lambda \left(h^2 -\alpha^2 \sigma^2\right)^2,
\end{align}
with $\lambda$ and $\alpha$ dimensionless and $h$ a scalar field. After letting $\sigma=\langle \sigma\rangle + \delta\sigma$, this generates a mass term for $h$ with mass $m^2=4\lambda \alpha^2\langle \sigma\rangle^2$. Note that this potential has a flat direction $\langle h\rangle=\alpha \langle\sigma\rangle$ which spans a one-dimensional manifold of vacua.

In principle we are not restricted to use a dilaton to make this trick. We could replace dimensionful quantities by powers of \emph{any operator} ${\cal O}$ that satisfies two conditions. 1) It has scaling dimension $d_{\cal O}=1$ and 2) It develops a non-vanishing vev. However, for the sake of simplicity and as a \emph{proof of concept}, we will stick here to the case where ${\cal O}\equiv\sigma$, where the dilaton may equally be a background field or one of the components on the multiplet $\phi^ a$.

This idea does not translate to QFT though, due to the presence of scale anomalies. Loop computations generically contain divergences which require the introduction of an external scale, in the form of a cut-off or reference scale, to renormalize the theory. The explicit presence of this quantity breaks scale invariance. This can be understood at the path integral level, as suggested by Fujikawa \cite{Fujikawa:1980vr}. Let us define the Euclidean partition function
\begin{align}
{\cal Z}[J]=\int [d\phi]e^{-S[\phi]-J\cdot \phi},
\end{align}
where we have introduced a source $J_ a$ and defined
\begin{align}
J\cdot \phi=\int d^4x \ J_a \phi^a.
\end{align}

Under scale transformations, assuming that the action is invariant and that the source is defined as an operator with the right scaling dimension $d_{J_a}=-d_{\phi^ a}$, the path integral transforms as
\begin{align}
\delta {\cal Z}[J]=\int (\delta [d\phi])e^{-S[\phi]-J\cdot \phi},
\end{align}
and therefore all non-invariance of ${\cal Z}[J]$ is induced by that of the functional integration measure. Normally, the explicit form of this measure is ignored, since it does not contribute to correlators in common settings \cite{Fujikawa:1983im,Fradkin:1974df,Toms:1986sh}. However, it is a key element in our construction here. We define it by following the conventions of \cite{Falls:2017cze,Falls:2018olk}
\begin{align}
[d\phi]=\prod_{a,x} \frac{d\phi^a(x)}{\sqrt{2\pi}}\sqrt{\det C}\ .
\end{align}
Here the product runs over all elements in $a$ and over all space-time points, while the factor $\sqrt{2\pi}$ is introduced for normalization purposes. $C_{ab}(x,y)$ is a bi-linear acting as a metric in field space and defines the following scalar product
\begin{align}
\{\phi^a (x),\phi^ b(y)\}=\int d^4 x \int d^4 y \ \phi^a(x) C_{ab}(x,y)\phi^b(y).
\end{align}
From now on we will restrict ourselves to ultra-local metrics in field space, given by
\begin{align}
C_{ab}(x,y)=s(x)\delta_{ab} \delta^{(4)}(x-y),
\end{align}
with all non-trivial dependence on space-time contained in the scalar function $s(x)$. 

The role of $C_{ab}$ in the quantization of the theory can be understood by looking at the one-loop correction for a given action $S[\phi]$. Expanding the fields as $\phi^a=\bar{\phi}^a + \delta \phi^a$ and using the standard background field method \cite{Abbott:1981ke,Barvinsky:2017zlx}, the one-loop partition function can be written in a closed form
\begin{align}
{\cal Z}^{(1)}[J]=e^{-S(\bar{\phi})}e^{\frac{1}{4}J_a (D^{-1})^{ab}J_b}\ \det\left( \frac{D_{ab}(x,y)}{s(x)}\right)^{-\frac{1}{2}},
\end{align}
where $D_{ab}(x,y)$ is the Hessian of the action and we have assumed that $\bar{\phi}$ is a solution of the classical equations of motion.

By looking at this formula we can see that if $s(x)=\Lambda^2$ this would correspond to the usual cut-off regularized path integral. Therefore, we recognise the role of the field space metric as that of a regulator. Physically, it puts ultra-violet contributions, corresponding to covering very large distances in field space, at a finite distance that can be now manipulated. Choosing different functions $s(x)$ thus corresponds to different regularization schemes for the QFT.

The transformation of the path integral measure under dilatations can now be easily obtained by using the scaling properties of the delta function. Under an infinitesimal change $\Omega=1+\omega+{\cal O}(\omega^2)$ we have
\begin{align}\label{eq:transf_measure}
\delta [d\phi]=\frac{(2-d_s)\omega}{2}{\rm Tr}\left(\delta_a^b \delta^{(4)}(x-y)\right) [d\phi].
\end{align}
where $d_s$ is the scaling dimension of $s(x)$. Formally, the trace in this formula is divergent. We will come back later to the problem of regularizing it, but for now we will just carry on forward keeping it as a formal expression. From \eqref{eq:transf_measure} we can see that, whenever the scaling dimension of $s(x)$ is $d_s\neq 2$, the path integral will not be SI. This is indeed the case when $s(x)\equiv\Lambda^2$ and we can recognize in \eqref{eq:transf_measure} the formula of Fujikawa \cite{Fujikawa:1980vr} for the scale anomaly. 

The advantage of using the integration measure as a regulator is that we are not tied to a particular perturbative computation nor to a given loop order at the time of introducing a regularization scheme. Indeed, we can now solve the problem of the scale anomaly by borrowing the idea of SIR, generalizing it to all orders in the path integral. We thus choose the regulator $s(x)$ to be the right power of the dilaton field $s(x)=\sigma^2$. With this $\delta[d\phi]=\delta {\cal Z}=0$ and we have
\begin{align}\label{eq:path_SI}
 {\mathfrak Z}[J]=\int\prod_{a,x} \frac{d\phi^a(x)}{\sqrt{2\pi}}\sqrt{\det\left( \sigma^2 \delta_a^b \delta^{(4)}(x-y)\right)}\ e^{-S[\phi]-J\cdot \phi},
\end{align}
where we have introduced the symbol ${\mathfrak Z}[J]$ to indicate the SI version of the partition function. As before, we are not forced to choose $s(x)=\sigma^2$ but we could use any operator ${\cal O}$ with the right scaling dimension. However, we stick here to this choice for the sake of simplicity.

This expression can be interpreted as a non-perturbative definition of SIR from first principles. The partition function \eqref{eq:path_SI} is exactly SI and valid up to arbitrary loop order as well as beyond perturbation theory. Note however than unlike in usual settings, here we have an explicit presence of a field in the integration measure. This poses a problem, since we need to take into account the contribution of the determinant when computing the path integral. 
\\
\paragraph*{{\bf Resolving the Path Integral} -}
In order to be able to compute correlators as well as to connect our top-down derivation with the bottom-up approach of SIR, we will make use of perturbation theory. On top of that, in order to be able to use standard techniques in computing quantum corrections, let us perform the following manipulation of the determinant
\begin{align}
\nonumber &\det\left( \sigma^2 \delta_a^b \delta^{(4)}(x-y)\right)\\
=&\det\left( \Lambda^2 \delta_a^b \delta^{(4)}(x-y)\right)\cdot\det\left( \frac{\sigma^2}{\Lambda^2} \delta_a^b \delta^{(4)}(x-y)\right),
\end{align}
so that we can write
\begin{align}
[d\phi]=(d\phi)\sqrt{\det\left( \frac{\sigma^2}{\Lambda^2} \delta_a^b \delta^{(4)}(x-y)\right)},
\end{align}
where we have defined the standard cut-off regularized measure
\begin{align}
(d\phi)=\prod_{a,x} \frac{d\phi^a(x)}{\sqrt{2\pi}}\sqrt{\det\left( \Lambda^2 \delta_a^b \delta^{(4)}(x-y)\right)}.
\end{align}

Using this, we can rewrite our SI partition function in a more traditional form
\begin{align}\label{eq:path_SI_2}
{\mathfrak Z}[J]=\int (d\phi)\sqrt{\det\left( \frac{\sigma^2}{\Lambda^2} \delta_a^b \delta^{(4)}(x-y)\right)}\ e^{-S[\phi]-J\cdot \phi},
\end{align}
which looks like a standard path integral, albeit with an extra determinant nested in.

The next step is to make this determinant explicit in terms of the fields which we are integrating over. We do that by using that
\begin{align}
\sqrt{\det\left( \frac{\sigma^2}{\Lambda^2} \delta_a^b \delta^{(4)}(x-y)\right)}=e^{\cal M},
\end{align}
where
\begin{align}
{\cal M}=\frac{1}{2}{\rm Tr}\log \left(\frac{\sigma^2}{\Lambda^2}\delta^{(4)}(x-y)\right).
\end{align}

Following \cite{Fujikawa:1980vr} we now introduce a basis of states of the theory $|\phi^a(x)\rangle$ in position space. These are defined as eigenvectors to the effective equations of motion regarded as operators
\begin{align}
\frac{\delta \mathfrak{L}}{\delta \phi^a(x)}\left[\ | \phi^a (x) \rangle\right]={\cal E}(x)\left[\ | \phi^a (x) \rangle\right]=\lambda_\phi |\phi^a(x)\rangle,
\end{align}
with $\lambda_\phi$ the corresponding eigenvalue. Here we are using $\mathfrak{L}[\phi]$ to denote the SI quantum effective Lagrangian obtained from $\mathfrak{Z}[J]$. Its integral over space-time defines the SI Quantum Effective Action $\mathfrak{F}[\phi]$. Using this we can write
\begin{align}
\nonumber {\cal M}&=\frac{1}{2}\int d^4x \langle \phi^a(x)| \log\left(\sigma^2/\Lambda^2\right)|\phi^a(x)\rangle\\
&=\frac{1}{2}\int d^4x \log\left(\sigma^2/\Lambda^2\right) \langle \phi^a(x)|\phi^a(x)\rangle,
\end{align}
where the repeated index denotes summation.

We are thus left with the task of computing the trace of the identity matrix, corresponding to summing over all eigenstates of the theory. Since this trace is divergent, we need to regularize it in some manner. We do it here by letting
\begin{align}
|\phi^a(x)\rangle=\lim_{M\rightarrow\infty}e^{-\frac{\lambda_{\phi}}{M^2}}|\phi^a(x)\rangle=\lim_{M\rightarrow\infty}e^{-\frac{{\cal E}(x)}{M^2}}|\phi^a(x)\rangle,
\end{align}
and therefore
\begin{align}
{\cal M}=\lim_{M\rightarrow \infty}\frac{1}{2}\int d^4x \log\left(\sigma^2/\Lambda^2\right) \langle \phi^ a(x)|e^{-\frac{{\cal E}(x)}{M^2}}|\phi^a(x)\rangle.
\end{align}

Plugging the completeness relation in momentum space and operating, we can finally write
\begin{align}
{\cal M}=\lim_{M\rightarrow \infty}\frac{1}{2}\int d^4x \log\left(\sigma^2/\Lambda^2\right) \int \frac{d^4k}{(2\pi)^4}e^{-\frac{{\cal E}(k)}{M^2}},
\end{align}
where the exponent must be now evaluated in momentum space.

Up to this point it looks like we have not gained anything. In order to compute the path integral we need the value of ${\cal M}$, which depends upon the path integral itself due to the explicit presence of $\mathfrak{L}[\phi]$. However, things become manageable in perturbation theory. Let us then assume that we can expand the classical action and thus the quantum effective action in powers of a coupling $g\ll 1$
\begin{align}
&S[\phi]=S_0[\phi]+g S_{\rm int}[\phi],\\
&\mathfrak{L}[\phi]={\cal L}_0[\phi]+g {\cal L}_{\rm int}[\phi]+\sum_{n=2}g^2 \mathfrak{L}_n[\phi],
\end{align}
where ${\cal L}_0[\phi]$ and ${\cal L}_{\rm int}[\phi]$ refer to the free and interaction parts of the Lagrangian respectively, while $\mathfrak{L}_n[\phi]$ are the successive contributions coming from loop corrections.

Let us further assume that we are dealing with relativistic fields with dispersion relation $k^2=0$ in the free case. Since the classical action cannot contain any dimensionful coupling due to SI, the interaction Lagrangian must be momentum independent. Then, it is easy to check by direct integration that at leading order we have
\begin{align}
{\cal M}\sim \lim_{M\rightarrow \infty} M^4\left( {\cal M}_0+\sum_{l=1} \left(\frac{g}{M^2}\right)^l  {\cal M}_l\right),
\end{align}
where ${\cal M}_i$ are functions of the fields that must be computed case by case. Although this result has divergent terms when we remove the cut-off $M$, they can be absorbed into a shift of the integration measure and the couplings of the classical theory.

Once the leading order of the expansion has been determined, we can expect that at an arbitrary loop order we will have
\begin{align}
{\cal M}=\sum_{l=2} g^l {\cal M}_l,
\end{align}
where the $g^2$ term can be computed by using solely the classical Lagrangian, while higher orders require to account for loop corrections. This is actually a huge advantage, since it means that the contribution from the determinant to the vertices starts at next-to-leading order.

Collecting this knowledge, we can formally write a perturbative expansion of $\mathfrak{Z}[J]$ as
\begin{align}\label{eq:pert_path}
{\mathfrak Z}[J]=\int (d\phi)\ e^{-S_0[\phi]-g S_{\rm int}[\phi]-J\cdot \phi+\sum_{l=2} g^l {\cal M}_l},
\end{align}
which can now be iterated and computed order by order in powers of $g$.

This perturbative procedure exactly matches SIR. In there, one first computes the one-loop correction to the effective action in a standard setting. Afterwards, they substitute the cut-off -- or equivalently the renormalization scale $\mu$ in dimensional regularization -- by the dilaton $\sigma$. Finally, they expand the result and this generates new vertices that must be taken into account for computations at higher loops. 

In our language, we would first compute the one loop correction coming from \eqref{eq:pert_path}. This includes two pieces, the standard one coming from computing using the classical action, and the correction given by the determinant. When taken together, they combine into exactly the SI result found by SIR, without need to substitute $\Lambda$ in an adhoc way. Furthermore, expression \eqref{eq:pert_path} already makes clear that the determinant must contribute with new vertices for higher order corrections. Note that this will potentially make the theory non-renormalizable, even if the starting action was renormalizable with standard quantization techniques.
\\
\paragraph*{{\bf An example: $\mathbf{\lambda\phi^4}$} -}
Although the procedure just described is clear and concise, it might be cumbersome when described in an abstract way. In order to clarify how to perform the computations, let us work out a simple example, that of a scalar field with a $\lambda \phi^4$ interaction at one-loop. We thus introduce the action
\begin{align}
S[\phi]=\int d^4x \left(\frac{1}{2}\partial_\m\phi \partial^\m \phi + \frac{\lambda}{4!}\phi^4\right),
\end{align}
where we are already using Euclidean signature, thus the sign of the potential term. We assume that the dilaton $\sigma$ is a background field in this case.

Let us start by computing the contribution ${\cal M}$ of the determinant. At the leading order, the effective equations of motion reduce to the classical ones
\begin{align}
\left(-\square +\frac{\lambda}{6}\phi^2\right)\phi=0,
\end{align}
and therefore we have
\begin{widetext}
\begin{align}
{\cal M}= \frac{1}{2}\int d^4 x\log\left(\sigma^2/\Lambda^2\right)\int \frac{d^4k}{(2\pi)^4}\exp\left(-\frac{k^2}{M^2}-\frac{\lambda}{6 M^2}\phi^2\right)=\frac{1}{4\pi^2}\int d^4x \log\left(\sigma^2/\Lambda^2\right)\left(M^4 - \frac{\lambda M^2}{6}\phi^2 +\frac{\lambda^2}{36}\phi^4\right),
\end{align}
\end{widetext}
where we have omitted the $\lim$ symbol and we have performed an expansion in powers of $\lambda$, ignoring terms which vanish when $M\rightarrow \infty$. Absorbing the divergences and renormalizing, we are left with a finite piece
\begin{align}
{\cal M}=\frac{\lambda^2}{144\pi^2}\int d^4x\ \log\left(\sigma^2/\Lambda^2\right)\phi^4.
\end{align}

Now we go back to the path integral. Since ${\cal M}$ is already order $\lambda^2$, the vertices obtained from it will not contribute at this order. Thus, we can take it out of the integral and write
\begin{align}
\mathfrak{Z}[J]=e^{\cal M}\int (d\phi)e^{-S[\phi]-J\cdot \phi}=e^{\cal M}Z[J],
\end{align}
where we have introduced back the non-SI integral $Z[J]$. If we know define the SI effective action $\mathfrak{F}[\phi]$ by taking a Legendre transform of  ${\cal W}[J]=-\log \mathfrak{Z}[J]$ as usual, we get
\begin{align}
\mathfrak{F}[\phi]=\Gamma[\phi]-{\cal M},
\end{align}
where $\Gamma[\phi]$ is the standard Quantum Effection Action. If we furthermore focus on the effective potential, we have
\begin{align}
\mathfrak{V}[\phi]=\left.\mathfrak{F}[\phi]\right|_{p=0}=V_{\rm eff}[\phi]-\left.{\cal M}\right|_{p=0}.
\end{align}

The first term $V_{\rm eff}[\phi]$ is well-known for the case of the scalar field. It corresponds to the Coleman-Weinberg potential \cite{Coleman:1973jx} and reads
\begin{align}
V_{\rm eff}[\phi]=\frac{\lambda^2}{144 \pi^2}\int d^4x \log\left(\phi^2/\Lambda^2\right)\phi^4.
\end{align}

It is not an accident that $V_{\rm eff}[\phi]$ and ${\cal M}$ are almost identical. If we take into account that $\mathfrak{V}[\phi]$ must be SI by construction, we have that
\begin{align}
\frac{\partial V_{\rm eff}[\phi]}{\partial\Lambda}=\frac{\partial{\cal M}}{\partial\Lambda}.
\end{align}

We can actually recognize the lhs of this expression as the scale anomaly of the effective action, leaving us with the knowledge that ${\cal M}$ is the integral of the anomaly over the cut-off parameter. Moreover, note that due to the way that $\Lambda$ is introduced, we have $\partial{\cal M}/\partial\log\Lambda + \partial{\cal M}/\partial\log\sigma=0$. This allows us to exchange integration over $\Lambda$ by integration over $\sigma$. Then we can recognize ${\cal M}$ as a Wess-Zumino term that removes the anomaly when added to the effective action.

Putting together both contributions we find for the effective potential
\begin{align}
\mathfrak{V}[\phi]=\frac{\lambda^2}{144 \pi^2}\int d^4x \log\left(\phi^2/\sigma^2\right)\phi^4.
\end{align}

As expected, this result is exactly scale invariant and coincides with that of \cite{Ghilencea:2016ckm} when comparisons are possible. If we were to perform now a two-loop computation, we would need to expand it around the vev of $\sigma$ and account for new non-renormalizable interactions coming from the logarithm, corresponding to the terms ${\cal M}_l$ with $l\geq 2$ in \eqref{eq:pert_path}.
\\
\paragraph*{{\bf Conclusions} -}
In this \emph{letter} we have presented a path integral formulation of SIR. We have done it by modifying the functional measure of integration, extending it with the presence of an extra determinant that cancels out the scale anomaly exactly. When expanded in perturbation theory, the new determinant plays a double role. It cancels the contribution of the anomaly at every order, and it also provides new interaction vertices that make the theory generically non-renormalizable. Our main results are thus expressions \eqref{eq:path_SI} and \eqref{eq:path_SI_2}. They provide a way to define SIR from first principles, but they also potentially allow for non-perturbative computations or even for application of lattice techniques.

Moreover, the SI path integral \eqref{eq:path_SI} represents a \emph{novel formulation} of exactly SI QFTs. It extends the known landscape of SI theories by allowing to formulate a SI QFT out of any classically SI theory. In particular, it allows to formulate theories where SI is spontaneously broken in the IR while exactly preserved by quantum effects.

Finally, let us point out that this formalism can be extended to allow for \emph{Weyl} invariance in the presence of Gravity, just by replacing the regulator $s(x)$ by the appropriate Weyl covariant object.
\\
\paragraph*{{\bf Acknowledgements} -}
I am grateful to K. Falls and S. Mooij for discussions. My work has been supported by the European Union's H2020 ERC Consolidator Grant “GRavity from Astrophysical to Microscopic Scales” grant agreement no. GRAMS-815673. I also wish to acknowledge networking support from COST action CA16104 ``GWverse".

%%%%%%%%%%%%%%%%%%%%%%%%%%%%%%%%%
%%%%%%%%%%%%%%%%%%%%%%%%%%%%%%%%%
\bibliography{scale}{}

%%%%%%%%%%%%%%%%%%%%%%%%%%%%%%%%%%%%%%%%%
\end{document}